\documentclass[twocolumn,prb,color,psfig,superscriptaddress]{revtex4}
\usepackage{graphicx}
\usepackage{epsfig}

\def\beq{\begin{equation}}
\def\eeq{\end{equation}}

\begin{document}

\title{What is the true charge transfer gap in parent insulating cuprates?}
\author{A.S.~Moskvin}
\affiliation{Ural State University, 620083 Ekaterinburg,  Russia}
\date{\today}


\begin{abstract}

A large body of experimental data point towards a charge transfer instability of parent insulating cuprates to be their unique property. We argue that the true charge transfer gap in these compounds is as small as 0.4-0.5\,eV rather than 1.5-2.0\,eV as usually derived from the optical gap measurements. In fact we deal with a competition of the conventional (3d$^9$) ground state and a charge transfer (CT) state with formation of electron-hole dimers which evolves under doping to an unconventional bosonic system. Our conjecture does provide an unified standpoint on the main experimental findings for parent cuprates including linear and nonlinear optical, Raman, photoemission, photoabsorption, and transport properties anyhow related with the CT excitations. In addition we suggest a scenario for the evolution of the CuO$_2$ planes in the CT unstable cuprates under a nonisovalent doping. 

\end{abstract}

\maketitle

\section{Introduction}

The origin of high-T$_c$ superconductivity\,\cite{Muller} is presently still a matter of great
controversy.
Copper oxides start out life as insulators in contrast with BCS
superconductors being conventional metals. Unconventional behavior of cuprates under charge doping, in particular, a remarkable interplay of charge, lattice, orbital, and spin degrees of freedom, strongly differs from that of ordinary metals and merely resembles that of a doped semiconductor.

We believe that unconventional behavior of cuprates can be consistently explained in frames of a so called dielectric scenario\,\cite{Moskvin} that implies their instability regarding  the \emph{d-d} charge transfer (CT) fluctuations.
Essential physics of the  doped cuprates, as well as many other
strongly correlated oxides, appears to be driven by a self-trapping of the CT excitons,
both one-, and two-center ones. Such excitons are the result of self-consistent charge transfer and lattice distortion with the appearance of a "negative-$U$" effect\,\cite{negative-U}.

At present, the CT instability  with regard to disproportionation is believed to be a rather
typical property for a number of perovskite 3d transition-metal oxides
 such as CaFeO$_3$, SrFeO$_3$, LaCuO$_3$, RNiO$_3$ \cite{Mizokawa}, RMnO$_3$\,\cite{LaMnO_3-pd} and  LaMn$_7$O$_{12}$\,\cite{LaMn_7O_12}, moreover,  in solid state chemistry  one consider tens of disproportionated systems\,\cite{Ionov}.  
 Phase diagram of disproportionated systems is
rather rich and incorporates different phase states from classical, or chemical disproportionated state to quantum states, in particular, to the unconventional  Bose-superfluid (superconducting) state\,\cite{Moskvin}.

Speaking about a close relation between disproportionation and superconductivity  it is worth noting a text-book example of BaBiO$_3$ system where we unexpectedly deal with the disproportionated Ba$^{3+}$+ Ba$^{5+}$ ground state instead of the conventional lattice of Ba$^{4+}$ cations\,\cite{BaBiO3}. The bismuthate can be converted to a superconductor by a nonisovalent substitution such as in Ba$_{1-x}$K$_x$BiO$_3$. At present,  this system seems to be the only one where the unconventional superconductivity is related anyhow with the disproportionation reaction. 

Regrettably physicists have paid remarkably little attention to the question of valence disproportionation and negative-U approaches ("chemical" route!) which are surely are being grossly neglected in all present formal theoretical treatments of HTSC.

The paper is organized as follows. In Sec.II we argue that the parent cuprates represent unconventional strongly correlated 3d oxides where strong electron-lattice polarization effects give rise to  an instability with regard to a charge transfer. In Sec.III we point to the midinfrared absorption universally observed in all the parent 2D cuprates to be a signature of the true charge transfer gap. In Sec.IV we address the structure and dispersion of the CT excitons, or electron-hole dimers in parent cuprates. In Sec.V we address different experimental data supporting our conjecture of anomalously small true CT gap in parent cuprates. In Sec.VI we discuss the evolution of the CT unstable parent cuprates under a nonisovalent doping.

\section{Electron-lattice relaxation and CT instability of parent cuprates} 

Minimal energy cost of the optically excited
disproportionation or electron-hole formation due to a direct Franck-Condon (FC) CT transition in insulating cuprates  is
$E_{gap}^{opt}\approx $1.5-2\,eV. This relatively small value of the
optical gap is addressed to be an argument against
the "negative-$U$" disproportionation reaction 2Cu(II) = Cu(III) + Cu(I)\,\cite{Good}, or
more correctly
\begin{equation}
\mbox{CuO}_{4}^{6-}+\mbox{CuO}_{4}^{6-}\rightarrow
\mbox{CuO}_{4}^{7-}+ \mbox{CuO}_{4}^{5-}\, .
\label{dispro}
\end{equation}
However, the question arises, what is the energy cost for the thermal
excitation of such a local disproportionation?  The answer implies
first of all the knowledge of relaxation energy, or  the energy gain due to the lattice polarization by the
localized charges. The full polarization energy $R$  includes the cumulative
effect of $electronic$ and $ionic$ terms, related with the displacement of
electron shells and ionic cores, respectively. The former term $R_{opt}$ is due
to the {\it non-retarded} effect of the electronic polarization by the momentarily
localized electron-hole pair given the ionic cores fixed at their perfect crystal positions.
Such a situation is typical for lattice response accompanying the FC
transitions (optical excitation, photoionization). On the other hand, all the
long-lived excitations, i.e., all the intrinsic thermally activated states and
the extrinsic particles produced as a result of doping, injection or optical
pumping should be regarded as stationary states of a system with a deformed
lattice structure. Thorough calculation of the localisation energy for the electron-hole pairs (EH-dimers) remains a challenging task for future studies. It is worth noting that despite their very large, several eV magnitudes, the relaxation  effects are not incorporated into current theoretical models of cuprates.

\begin{figure}[t]
\includegraphics[width=8.5cm,angle=0]{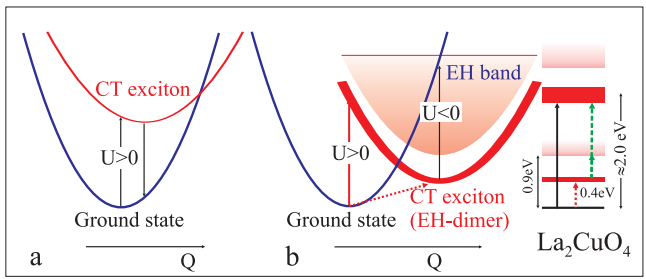}
\caption{(Color online) Simple illustration of the electron-lattice polarization effects for the CT excitons (see text for details): a) CT stable system; b) CT unstable system. Filling points to a continuum of unbounded electrons and holes. Right panel shows the experimentally deduced energy scheme for the CT states in La$_2$CuO$_4$, arrows point to various CT transitions} 
\label{fig1}
\end{figure}

Figure\,\ref{fig1} does illustrate two possible ways  the electron-lattice polarization governs the CT excitation evolution. Shown are the adiabatic potentials for the two-center ground state (GS) $M^0-M^0$ configuration and  excited $M^{\pm}-M^{\mp}$ CT, or disproportionated configuration.
A configurational coordinate $Q$ is associated with a lattice degree of freedom such as a half-breathing mode.
  For lower branch of adiabatic potential (AP) in the system we have either a single minimum point for the GS configuration (Fig.\,\ref{fig1}a) or  a two-well structure with an additional local minimum point (Fig.\,\ref{fig1}b) associated with the self-trapped CT exciton. This   "bistability" effect is of primary importance for our analysis. Indeed, these two minima are related with two (meta)stable charge states with and without CT, respectively,  which form two candidates to struggle for a ground state.  It is worth noting that the self-trapped CT exciton may be described as a configuration with  negative disproportionation energy $U$. 
 Thus one  concludes that all the systems such as 3d oxides may be separated to two classes: {\it CT stable systems} with the only lower AP branch  minimum for a certain  charge configuration, and bistable, or {\it CT unstable systems} with two lower AP branch minima for two local charge configurations one of which is associated with the self-trapped CT excitons  resulting from self-consistent charge transfer and electron-lattice relaxation.  

A large body of experimental findings point to an instability of the parent cuprates with regard  to a charge transfer (see, e.g., Ref.\onlinecite{Gorkov} and references therein). Maybe the most exciting evidence is obtained by the ultrafast electron crystallography (UEC) which does provide, through observation of spatiotemporally resolved diffraction, an unique tool for determining structural dynamics and the role of electron-lattice interaction\,\cite{Gedik}. A polarized femtosecond ($fs$) laser pulse excites the charge carriers, which relax through electron-electron and electron-phonon couplings, and the consequential structural distortion is followed diffracting $fs$ electron pulses. The technique has revealed a structural instability in La$_2$CuO$_4$ related with the CT excitations, or CT fluctuations\,\cite{Gedik}. Above a certain threshold, a direct conversion between two phases with distinct electronic and structural properties of the lattice  was observed, indicating that macroscopic scale domains (which define the coherence length of the Bragg diffraction) are involved in this phase
transformation. Thus the CT excitation,  during its
thermalization, induced distinct structural changes which distorted the lattice in a way that was observable at longer
times $\geq$\,300\,ps\,\cite{Gedik}. The very slow time scale reflected the fact that electronic and structural relaxations are coupled. In order for the charges to fully recombine, the lattice has to relax as well, which naturally takes a long time especially if the acoustic phonons are involved. 

\section{MIR band as a signature of the true CT gap in parent cuprates}

\begin{figure}[t]
\includegraphics[width=8.5cm,angle=0]{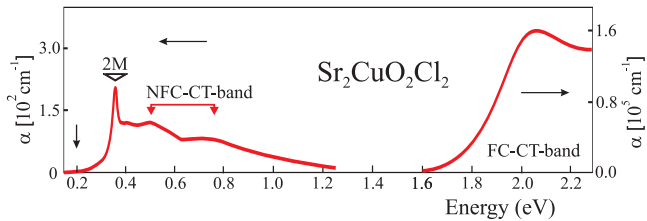}
\caption{(Color online) Reconstruction of the whole NFC-FC CT band in Sr$_2$CuO$_2$Cl$_2$. Low-energy MIR band is reproduced from Ref.\,\protect\onlinecite{Perkins}, the main FC CT band is taken from Ref.\,\protect\onlinecite{Cooper}. The scale of the respective absorption coefficients differs by three orders of magnitude. Such an unconventional NFC-FC structure of the optical spectra is a typical one for all parent cuprates. Vertical arrow points to a hardly visible peak at $\approx$\,0.2\,eV\,\protect\cite{MIR}.} \label{fig2}
\end{figure}

Unfortunately, the experimental information regarding the relaxation energies for CT excitons in 3d oxides is scarcely available. 
Just recently, by measuring the Hall coefficient $R_H$ up to 1000\,K in La$_2$CuO$_4$ Ono {\it et al.}\cite{Ando} have estimated the energy gap over which the electron and hole charge carriers are thermally activated in parent cuprate La$_2$CuO$_4$  to be $\Delta_{CT}$=0.89\,eV.  True chemical potential jump between the hole- and electron-doped Y$_{0.38}$La$_{0.62}$Ba$_{1.74}$La$_{0.26}$Cu$_3$O$_y$ (YLBLCO) was measured\,\cite{YLBLCO} to be $\approx$\,0.8\,eV. These energies may be interpreted as the minimal ones needed to create uncoupled electron-hole pairs. Hence the minimal energy $E_{gap}^{CT}$ of  the local disproportionation reaction with the creation of the relaxed bounded electron-hole pair, or EH-dimer, can be substantially less than 0.8\,eV that points to a dramatic CT instability of parent cuprate, especially, if remind of 1.5-2.0\,eV to be a minimal energy of optical creation of CT exciton, or bound EH-pair. This difference between the true quasiparticle gap that determines the transport and thermodynamics and the optically measured CT gap has also been found in electron-doped materials\,\cite{Xiang}. In Nd$_2$CuO$_4$ the band gap, measured as the minimum excitation energy between the hole and electron bands, is estimated to be only 0.5\,eV, much lower than the optically measured CT gap, which is usually believed to be about 1.5\,eV. In other words, the optical CT gap, which is generally deduced from the peak energy of the FC optical absorption, does not correspond to the true gap between the two bands in parent cuprates. 
However, the true CT gap $E_{gap}^{CT}$ would be optically detected as a low-energy edge of the weak non-FC (NFC) CT bands.
Indeed, the dipole matrix elements for a direct FC and a non-direct NFC CT transitions differ mainly because of different vibrational overlap integrals, big for the former and small for the latter. Obviously, the mid-infrared (MIR) band universally found in all parent cuprates\,\cite{Kastner,Perkins,Gruninger} extending from 0.4 up to 1\,eV  results mainly from the weak non-FC CT optical transition which final state corresponds to the low-energy  relaxing EH-dimers. In other words, the MIR band in the undoped cuprates is believed to present a non-FC counterpart of the main low-energy FC CT band. The whole  lineshape of the NFC+FC CT band shown in Fig.\,\ref{fig2} for Sr$_2$CuO$_2$Cl$_2$\,\cite{Perkins,Cooper,MIR} is typical for other parent cuprates and would strongly deviate from that of typical for the CT stable system. In particular, in the isostructural to La$_2$CuO$_4$ oxide La$_2$NiO$_4$ no such bands are observed\,\cite{Kastner} and the lineshape of the  MIR absorption band  in this antiferromagnet is perfectly consistent with the predictions of the purely spin model\,\cite{LS}.


Making use of experimental data\,\cite{Kastner,Perkins,Gruninger} we conclude that the true CT gap $E_{gap}^{CT}$ for parent cuprates such as La$_2$CuO$_4$, Nd$_2$CuO$_4$, Pr$_2$CuO$_4$, Sr$_2$CuO$_2$Cl$_2$, and YBa$_2$Cu$_3$O$_6$ is as small as 0.4-0.5\,eV. This puzzling result points to a remarkable charge transfer instability of parent cuprates. It also means that the charge fluctuations in high-T$_c$ materials are much stronger than usually believed and should be fully considered in the construction of the basic model of high-T$_c$ superconductivity. It is worth noting that the \emph{d-d} CT energy defines an effective $U_d$ parameter, hence its value in parent cuprates can be as small as 0.4\,eV.

It should be noted that the MIR band in parent cuprates is composed of a sharp lowest energy resonance peak and a clearly resolved two-peak high-energy structure\,\cite{Kastner} with  peaks near 0.4-0.5\,eV and 0.7-0.8\,eV, respectively (see Fig.\,\ref{fig2}). As for its FC counterpart\,\cite{Moskvin1,Moskvin2} we can relate these two peaks with  two-center \emph{d-d} ($b_{1g}\rightarrow b_{1g}$) and one-center \emph{p-d} allowed electro-dipole $b_{1g}\rightarrow e_u(\pi)$ NFC CT transitions, respectively. The electron-hole pair related with the latter transition is composed of the $b_{1g}\propto d_{x^2-y^2}$ electron and purely oxygen $e_u(\pi)$ hole. It is worth noting that the low-energy multiplet of the EH dimers can incorporate the $b_{1g}-\underline{a}_{2g}(\pi)$ pair composed of the $b_{1g}$ electron and purely oxygen $a_{2g}(\pi)$ hole. However, corresponding one-center \emph{p-d} CT transition is electro-dipole forbidden hence this NFC excitation as well as its 
FC counterpart\,\cite{A_2g-Raman} can be revealed only by Raman scattering technique.

The  energy of the lowest CT excitation is close to that of the two-magnon (2M) excitation obtained by flipping two spins on neighboring sites at the energy estimated in a spin-wave approximation as $E_{2M}=2.73\,J\approx$\,0.3-0.4\,eV\,\cite{Kastner}. It means their strong coupling with formation of a low-energy predominantly two-magnon excitation (main resonance peak at $\approx$\,0.35\,eV in  Sr$_2$CuO$_2$Cl$_2$ as seen in Fig.\,\ref{fig2})  and a high-energy predominantly CT excitation. The coupling reduces the energy of the 2M excitation that results in  lower values of exchange integrals calculated from the MIR peak position as compared with those found by neutron and Raman scattering\,\cite{Perkins,Gruninger}. In addition the coupling makes the 2M excitation partly dipole-allowed.

Strong coupling of the low-energy CT excitations with the high-energy magnetic excitations can explain strong deviations from the predictions of the spin wave theory observed recently by inelastic neutron scattering in the parent cuprate  La$_2$CuO$_4$\,\cite{SWT-124} at high energies near the top of the spin-wave band ($\approx$\,300\,meV).
While the lower energy excitations are well described by spin-wave theory, including one- and two-magnon scattering processes, the high-energy spin waves are strongly damped near the ($\pi ,0$) position in reciprocal space and merge
into a momentum dependent continuum. This anomalous damping indicates the decay of spin waves into
other excitations, possibly dispersive EH dimers.


The nature of the MIR band unique to the layered insulating cuprates remains to be
one of the old mysteries of the cuprate physics.  To explain experimental data
available for parent insulating cuprates Lorenzana and Sawatzky\,\cite{LS} (LS)
proposed mechanism of the phonon-assisted multimagnon absorption. The LS
mechanism allows successful interpretation of experimental data on MIR
absorption for $S=1$ 2D  antiferromagnet La$_2$NiO$_4$ but
fails to explain all the features of MIR in  the $s=1/2$ 2D cuprates La$_2$CuO$_4$, Nd$_2$CuO$_4$, Sr$_2$CuO$_2$Cl$_2$,
YBa$_2$Cu$_3$O$_6$\,\cite{Perkins,Gruninger,Struzhkin} except the lowest energy sharp resonance peak.
The intrinsic width and the line shape of the whole MIR band remain beyond a description in terms of a spin-only Hamiltonian and point to different physics. It is worth noting that the  mid-IR features are not  susceptible to external magnetic field. Within the error bars of the experiments, there is no systematic magnetic field-induced changes of the mid-IR transmission in La$_2$CuO$_4$ in magnetic field of 18\,T\,\cite{field-effect}. 


\section{EH-dimers in parent cuprates}

The two-center \emph{d-d} CT  excitons or electron-hole (EH-) dimers may be considered as quanta of the disproportionation
reaction (\ref{dispro})
with the creation of electron  CuO$_{4}^{7-}$ and hole
CuO$_{4}^{5-}$ centers. The former corresponds to completely
filled Cu $3d$ and O $2p$ shells, or the vacuum state for holes $|0\rangle$,
while the latter may be found in different two-hole states $|2\rangle$, first the ground state Zhang-Rice singlet\,\cite{ZR}. 
The two EH-dimers will interact due to a resonance reaction $|02\rangle \leftrightarrow |20\rangle$:
\begin{equation}
\mbox{CuO}_{4}^{7-}+\mbox{CuO}_{4}^{5-} \leftrightarrow
\mbox{CuO}_{4}^{5-} +\mbox{CuO}_{4}^{7-} \; ,
\label{eh-he}
\end{equation}
governed by an effective resonance two-particle (bosonic!) transfer integral $t_{eh}=t_B$.

The energies of the two respective superposition states 
$$
|\pm\rangle = \frac{1}{\sqrt{2}}(|02\rangle \pm |20\rangle)\, ,
$$
are given by $E_0 \pm |t_{B}|$, where $E_0$ is the energy of the
bare $|20\rangle$, $|02\rangle$ states. The even (odd)-parity states $|\pm\rangle $
correspond to $S$- or $P$-like two-center excitons. 
Let us note that in our approach the $S$- and $P$-excitons are
centered at the central oxygen ion of the Cu$_2$O$_7$ cluster shared by the both electron and hole centers.

The resonance reaction (\ref{eh-he}) corresponds to an inter-center transfer of {\it two holes}, or {\it two electrons}.
The magnitude of the effective resonance transfer
integral $t_{B}$ which determines both the excitonic even-odd or $S-P$ splitting and the two-particle transport is believed to be of a particular interest in cuprate physics.
It can be written as follows:
$$
t_{B}=  \langle 20| V_{ee}| 02\rangle - \sum_{11}\frac{\langle 20| \hat h| 11\rangle  \langle 11|  \hat h| 02\rangle }{\Delta_{dd}}\, ,
$$
where the first term describes a simultaneous tunnel transfer of the electron
pair due to Coulomb coupling $V_{ee}$ and may be called as "potential"
contribution, whereas the second describes a two-step (20-11-02) electron-pair transfer
via   successive one-electron transfer due to one-electron Hamiltonian  $\hat
h$, and may be called as "kinetic" contribution. As it is emphasized by P.W.
Anderson\,\cite{PWA} the value of the seemingly leading kinetic contribution to pair (boson!)
transport is closely related to the respective contribution to the exchange
integral, i.e. $t_{B}\approx$\,0.1\,eV.

The $S$-exciton is dipole-forbidden, in contrast to the
$P$-exciton, and corresponds to a so-called two-photon state.
However, these two excitons have a very strong dipole-coupling
with a large value of the $S$-$P$ transition dipole matrix
element:
\begin{equation}
\label{me} d = |\langle S|\hat{\bf d}|P\rangle |\approx 2eR_{CuCu}\approx 8e\AA\, .
\end{equation}
 This points to a very important role played by this
doublet in nonlinear optics, in particular in two-photon
absorption and third-harmonic generation effects\,\cite{Ogasawara,THG}. Indeed, the quasi-1D insulating chain cuprates Sr$_2$CuO$_3$ and Ca$_2$CuO$_3$ with corner shared CuO$_4$ centers  show anomalously large third-order optical nonlinearities as revealed by electroreflectance\cite{Kishida,Ono-04}, third-harmonic generation\cite{Kishida-01}, two-photon absorption\cite{Ogasawara,Maeda-04}. The model fitting of the nonlinear optical features observed near 2\,eV in Sr$_2$CuO$_3$ yields: $E_P=1.74$ eV, $E_S=1.92$ eV, $\langle S|x|P\rangle =10.5 \AA$\,\cite{Maeda-04} (or $\approx $ 8\AA\,\cite{Kishida}). Despite some discrepancies in different papers\cite{Kishida,Ono-04,Maeda-04} these  parameters agree both with theoretical expectations and the data obtained in other independent measurements. In other words, the nonlinear optical measurements provide a reliable estimation of the effective "length" of the two-center \emph{d-d} CT exciton and of the two-particle transfer integral: $t_B=\frac{1}{2}(E_S-E_P)\approx $\,0.1\,eV.

In the 2D case of an ideal CuO$_2$ layer we deal with two types of
$x$- ($S_{x},P_{x}$) and $y$- ($S_{y},P_{y}$) oriented $S,P$
excitons in every unit cell which dynamics in frame of the Heitler-London approximation
\cite{Davydov} could be described by an effective one-particle
excitonic Hamiltonian with a standard form as follows:
\begin{equation}
\hat H_{exc}=\sum _{\Gamma _{1}\Gamma _{2}{\bf R}_{1}{\bf R}_{2}}
\hat B^{\dagger}_{\Gamma _{1}}({\bf R}_{1}) T_{\Gamma _{1}\Gamma
_{2}}({\bf R}_{1}-{\bf R}_{2}) \hat B_{\Gamma _{2}} ({\bf R}_{2})
\end{equation}
in a site representation with $\hat B^{\dagger}_{\Gamma _{1}}({\bf
R}_{1})/ \hat B_{\Gamma _{2}} ({\bf R}_{2})$ being the excitonic
creation/annihilation operators, or
\begin{equation}
\hat H_{exc}=\sum _{\Gamma _{1}\Gamma _{2}{\bf k}} \hat
B^{\dagger}_{\Gamma _{1}}({\bf k})T_{\Gamma _{1}:\Gamma _{2}}
 ({\bf k})\hat B_{\Gamma _{2}} ({\bf k})
\end{equation}
in ${\bf k}$ representation. Here the $\Gamma _{1,2}$ indices label
different $S$- or $P$-excitons.

The $T({\bf k})$ matrix for an
isolated quartet of $S_{x,y}$ and $P_{x,y}$ excitons in 2D cuprates
can be written  as follows\,\cite{Ng,condmat} 
\begin{widetext}
 \begin{equation}
T({\bf k})= \pmatrix{E_{S}+2T_{S}^{\parallel}\cos k_{x}&-2iT_{SP}^{\parallel}\sin k_{x}&
T_{S}^{\perp}(1+a(k_{x},k_{y}))& T_{SP}^{\perp }(1+b(k_{x},k_{y}))\cr
2iT_{SP}^{\parallel}\sin k_{x}&E_{P}+2T_{P}^{\parallel}\cos k_{x}&
T_{SP}^{\perp }(1-b(k_{x},k_{y}))& T_{P}^{\perp }(1-a(k_{x},k_{y}))\cr
T_{S}^{\perp }(1+a^{*}(k_{x},k_{y}))& T_{SP}^{\perp }(1-b^{*}(k_{x},k_{y}))&
E_{S}+2T_{S}^{\parallel}\cos k_{y}&-2iT_{SP}^{\parallel}\sin k_{y}\cr
T_{SP}^{\perp }(1+b^{*}(k_{x},k_{y}))& T_{P}^{\perp }(1-a^{*}(k_{x},k_{y}))&
2iT_{SP}^{\parallel}\sin k_{y}&E_{P}+2T_{P}^{\parallel}\cos k_{y}\cr}, \label{matrix}
\end{equation}
\end{widetext}
where $a(k_{x},k_{y})=e^{ik_{x}}+e^{-ik_{y}}$,
$b(k_{x},k_{y})=e^{ik_{x}}-e^{-ik_{y}}$. Two diagonal $2\times 2$
blocks in this matrix are related with $S_x ,P_x$ and  $S_y ,P_y$
excitons, respectively;  off-diagonal blocks describe its coupling.
Here we have introduced a set of
transfer parameters to describe the exciton dynamics:
$$
T_{S}^{\parallel} \approx  -T_{P}^{\parallel} \approx \frac{1}{2}(t_{e}^{(3)}+t_{h}^{(3)}) ;\,\,
T_{SP}^{\parallel}\approx  \frac{1}{2}(t_{e}^{(3)}-t_{h}^{(3)}) \; ,
$$
for the collinear exciton motion, and
$$
T_{S}^{\perp}\approx -T_{P}^{\perp} \approx
 \frac{1}{2}(t_{e}^{(2)}+t_{h}^{(2)}) \, ;\,\,
T_{SP}^{\perp}\approx  \frac{1}{2}(t_{e}^{(2)}-t_{h}^{(2)}) \, ,
$$
for the $90^{\circ}$ rotation of the exciton.
The $90^{\circ}$ rotation, or "crab-like" motion, and  $180^{\circ}$, or collinear motion  of the exciton are governed by the one-particle (electron/hole) transfer integrals $t_{e,h}^{(2,3)}$ 
 for the next- ($nnn$) and next-next- ($nnnn$) nearest CuO$_4$ centers, respectively.  Hereafter we neglected higher order terms which seem to be less important.
 

All these parameters have a rather clear physical sense. The electron (hole) transfer integrals $t_{e,h}^{(3)}$ for collinear exciton transfer  ($R_{nnn}\approx 8$\AA) are believed to be smaller  than
$t_{e,h}^{(2)}$ integrals for rectangular transfer ($R_{nnn}\approx 4\sqrt{2}$\AA). In other words,
the two-center excitons prefer to move "crab-like", rather than in
the usual collinear mode. This implies a large difference for the
excitonic dispersion in (0,0)-(0,$\pi$) and (0,0)-($\pi$,$\pi$) directions. The electronic
wave function in the exciton (contrary to the hole one) has a
dominant Cu $3d$ nature that implies a smaller value of the
$t_{e}^{(2,3)}$ parameters compared to the $t_{h}^{(2,3)}$ ones.

\begin{figure}[t]
\includegraphics[width=8.5cm,angle=0]{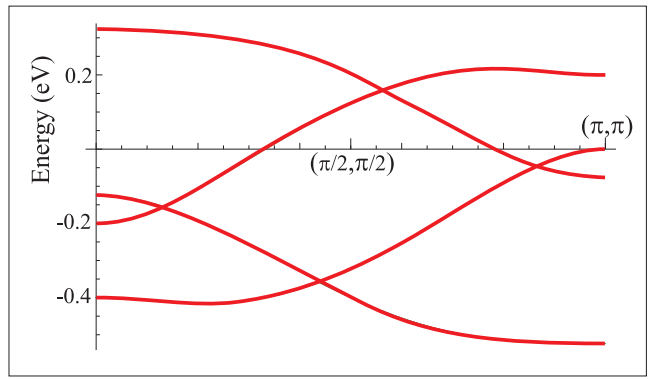}
\caption{(Color online) Dispersion of four EH-dimer modes in the nodal (0,0)-($\pi$,$\pi$) direction} \label{fig3}
\end{figure}

It is worth noting that in $\Gamma$ point (0,0) the excitons form four modes with the $A_{1g},B_{1g}$, and $E_u$ symmetry. In Fig.\ref{fig3} we present an example of the calculated excitonic dispersion along the nodal (0,0)-($\pi$,$\pi$) directions given reasonable values of different parameters: $(E_S-E_P)=2|t_B|=$\,0.2\,eV; $T_{S}^{\perp}=-T_{P}^{\perp}=T_{SP}^{\perp}=$\,0.1\,eV, $T_{S}^{\parallel}=T_{P}^{\parallel}=T_{SP}^{\parallel}=0$. In other words, we assume a nearest-neighbor approximation for the exciton transfer, and neglect the electron transfer integrals as compared with the hole ones.

 \section{Another experimental signatures of the anomalously small true CT gap in parent cuprates}

Despite a successful explanation of the MIR absorption band features our main conjecture needs in a further independent experimental validation. First of all it is worthwhile  to notice one remarkable optical feature which was overlooked in earlier measurements.  
 A weak but well defined peak at $E_0$\,=\,1570\,cm$^{-1}$ (195\,meV) in the optical conductivity has been observed recently in  Sr$_2$CuO$_2$Cl$_2$\,\cite{MIR}. The peak appears to strengthen and turn into a broad band with doping, whose peak softens rapidly. Such a behavior seems to be a typical one for the dipole-allowed S-P transition in the condensed EH-dimers which transforms into a broad bosonic band with doping. It is worth noting that a similar peak at $E_0\approx$\,1600\,cm$^{-1}$ is clearly seen in the optical conductivity spectra of YBa$_2$Cu$_3$O$_6$\,\cite{Gruninger}. These experimental findings provide an unique opportunity to estimate the numerical value of the two-particle, or local boson transfer integral $t_B$: $t_B\approx$\,0.1\,eV, that is the value we have obtained from the nonlinear optical measurements.
 
  
\subsection{Photoinduced absorption }
Low-energy metastable EH-dimers  can be detected by  photoinduced absorption (PA) measurements. The PA spectroscopy has become a very productive tool in the study  both of the ground and excited  electronic states. The energies and dynamics of the observed optical absorption are a sensitive tool to determine  the origins of the electronic energy gap within which these photoinduced absorptions are observed. 

\begin{figure}[t]
\includegraphics[width=8.5cm,angle=0]{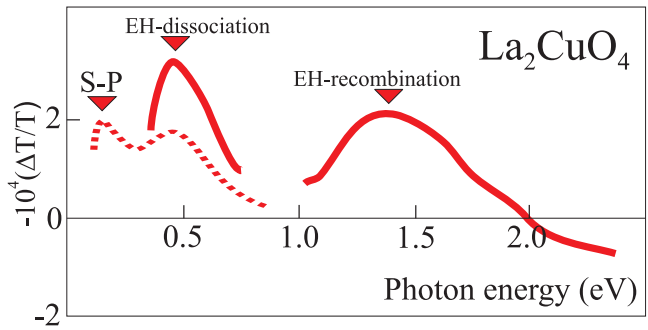}
\caption{(Color online) Photoinduced absorption spectrum of La$_2$CuO$_4$ at 15\,K taken with a pump photon energy of 2.54\,eV\,\protect\cite{Ginder}. Dotted curve presents the PA spectrum of La$_2$CuO$_4$ (arb.units) at 4.2\,K taken with a pump photon energy of 2.7\,eV\,\protect\cite{Kim}.}  \label{fig4}
\end{figure}

Two long-lived photoinduced absorption features  peaking at 0.5 and 1.4\,eV are observed in La$_2$CuO$_4$\,\cite{Ginder} with a crossover to photoinduced bleaching above 2.0\,eV (see Fig.\,\ref{fig4}). These data, together with observed luminescence at $\leq$\,2\,eV, confirm the existence of long-lived stable excited electronic CT states in this system.
The PA peak at 0.5\,eV can be naturally related with a photo-dissociation of the EH-dimers, while a high-energy PA peak at 1.4\,eV can be related with a photo-recombination of the EH-dimers, or inverse CT transition with the EH pair annihilation.
 
A little bit later the photoexcitation measurements for La$_2$CuO$_4$ and Nd$_2$CuO$_4$ 
by Kim {\it et al}.\,\cite{Kim} have revealed a more intricate structure of the low-energy photoinduced absorption band with two peaks at 0.12 and 0.47\,eV in La$_2$CuO$_4$ (see Fig.\,\ref{fig4}) and 0.16 and 0.62\,eV in Nd$_2$CuO$_4$ with 
additional bleaching of the in-plane phonon breathing modes. These low-energy peaks should be unambiguously  attributed to $S-P$ transitions in photo-generated EH-dimers.
Photoinduced absorption features  peaking near  1.5\,eV with a crossover to photoinduced bleaching near 2.0\,eV have been observed also in insulating  Nd$_2$CuO$_4$ and YBa$_2$Cu$_3$O$_{6.2}$\,\cite{Matsuda}. Similar effects have been recently observed in Sr$_2$CuO$_2$Cl$_2$\,\cite{PA-SCOC}.
All that does strongly support our scenario and energy scheme in Fig.\ref{fig1}. 

\subsection{Raman scattering spectroscopy}

Existence of low-energy CT excitations explains long-standing  troubles in the Raman scattering spectra of parent insulating and doped cuprates. Usually the Raman scattering process is described by an effective  Fleury-Loudon-Elliott spin Hamiltonian\,\cite{Raman}, which assumes the both initial and final states to lie well below the charge transfer gap.
However, as for the LS theory of MIR absorption, the spin only  theory of the Raman scattering runs into several difficulties. It cannot explain the large width with a clear asymmetry  extending towards high energies\,\cite{Raman}. 
The most notable discrepancy with the Fleury-Loudon-Elliott theory is that in addition to theoretically predicted $B_{1g}$ excitation at  in the experiments there is comparable scattering intensity in $A_{1g}$ polarizations and even in $A_{2g}$ and $B_{2g}$ polarizations of incident and outgoing light\,\cite{Raman}. 
However, an additional source for high-energy spectral features with the enhanced spectral weight and a complete collection of symmetries naturally arises from the coupling to the charge degrees of freedom. Indeed, at variance with the only $B_{1g}$ spin excitation our scenario implies existence of a whole collection of the CT excitations (EH-dimers) in the spectral range under consideration,  with the $E_u$, $A_{2g}$ symmetry for the \emph{p-d} CT transitions and $E_u$, $A_{1g}$, $B_{1g}$  symmetry for the \emph{d-d} CT transitions, embracing both electric-dipole-allowed and forbidden electronic excitations displaying itself in MIR absorption and Raman scattering, respectively.



\begin{figure}[t]
\includegraphics[width=8.5cm,angle=0]{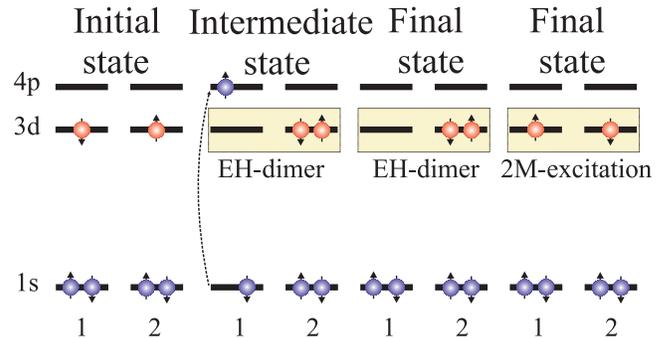}
\caption{(Color online) Possible resonant scattering process resulting in the
creation of the EH-dimer or the two-magnon excitation. The Cu (hole) spin on site 1 is
repelled onto a neighboring site, site 2, by the 1s core hole in
the intermediate state. Following the decay of the core-hole,
the "wrong" spin can hop back resulting in spin flips on both
sites.} \label{fig5}
\end{figure}

Direct observation of the low-energy \emph{d-d} CT transitions in Sr$_2$CuO$_2$Cl$_2$ has been performed in Ref.\onlinecite{RSXRS-SCOC}  using symmetry-selective resonant soft x-ray Raman scattering (RSXRS) experiments at the O 1s edge excitation. Taking advantage of extremely weak elastic scattering intensity in the O 1s edge the authors could observe  both the generic 2\,eV feature and a weak RSXRS structures around 0.5\,eV in the controversial midinfrared
region. The same photon polarization properties for the both bands points to their common nature.  In our opinion these are the FC and NFC \emph{d-d} CT bands, respectively. (The two-magnon excitation can in principle also be observed at the oxygen K edge (1s$\rightarrow $2p transition). Although the cross section for the oxygen K edge is relatively small, this case is interesting because the single magnon excitation is forbidden here: there is no spin-orbit coupling for 1s core orbitals.  Dispersive $\Delta S=0$ excitations have been observed in La$_2$CuO$_4$ both at the Cu K and L edges\,\cite{Hill,Braicovich}.  
Figure\,\ref{fig5} shows a possible RSXRS scattering mechanism. The system begins in the ground state, with
nearest neighbor 3d$^9$ spins antiferromagnetically coupled. A 1s core-level electron is then excited
into the 4p band. The resonance utilized in these experiments is that of the "well-screened" intermediate state, in which charge has moved in to screen the core-hole from the oxygen ligand state. Further, it is energetically favorable for this hole to form a Zhang-Rice singlet on the neighboring site thus creating a EH-dimer.  When the 4p decays, the "wrong" spin hole can hop back with the net effect of the flipping of the two spins.

\subsection{Photoemission spectroscopy}

Angle-resolved photoemission spectroscopy  (ARPES) is addressed to be a key
experiment to elucidate a number of the principal issues of electronic theory
related to the unconventional properties of cuprates\,\cite{Damascelli}.
Theoretically, ARPES measures the energy of out-coming photoelectron with a known energy and momentum with respect to the chemical potential's energy position. One supposes that ARPES can reconstruct the electronic band spectrum
of a system in the whole Brillouin zone. However, nature of the photoemission process itself, or in other words, the way the incident photon couples with the electronic states of the system in generating the photoemitted electrons is not yet understood. Even after years of intense ARPES studies for cuprates there is still no full understanding of the renormalization effects and of the relevant energy scales in their electronic excitation spectrum.

ARPES is the fast technique that implies an engagement of strong electron-lattice polarization effects that gives rise to a specific shape of the photoemission spectra which reflects  intensive FC transitions as well as weak NFC transitions. 
The whole  spectral weight associated with a certain electron-removal state for the CT unstable parent cuprates will be spread over energies as large as several eV's with a significant structure in momentum space. 
Obviously, these effects can hardly be catched by the simple t-J-Holstein model, where a photohole interacts with dispersionless optical phonons with the energies $<$\,0.1\,eV  through on-site local coupling\,\cite{Mishchenko}. Experimentally, usually one presents an ARPES study of the low binding-energy occupied electronic structure, which corresponds to an investigation of the low-energy  states. It is worth noting that {\it the true first electron-removal state certainly corresponds to the relaxed state, hence its ARPES portrait is formed by weak NFC transitions}. Such a situation makes the analysis of ARPES spectra to be  of a great ambiguity.

Photoemission process for parent insulating cuprate implies an overcoming of the true CT gap. In other words, one way or another, the photoemission process 
\begin{equation}
CuO_4^{6-}+h\nu \rightarrow	CuO_4^{5-}+e\, ,
\end{equation}
with creation of a free electron
should  start with  the excitation of the bound  electron-hole pair and the photohole is born due to a reaction as follows: 
\begin{equation}
CuO_4^{6-}+h\nu \rightarrow	\left[CuO_4^{6-}\right]^* \rightarrow	\left[CuO_4^{5-}\right]^*+e\, ,
\end{equation}
or
$$
CuO_4^{6-}+CuO_4^{6-}+h\nu \rightarrow	CuO_4^{5-}+CuO_4^{7-} \rightarrow
$$	
\begin{equation}
CuO_4^{5-}+CuO_4^{6-}+e\,
\end{equation}
for one- and two-center electron-hole pairs, respectively. Here, $\left[CuO_4^{6-}\right]^*$ denotes a \emph{p-d} CT state of the $CuO_4^{6-}$ center, while  $\left[CuO_4^{5-}\right]^*$ corresponds to a hole center in the low-energy non-ZR states with the nominal 3d$^{10}$ configuration on the Cu site.
Indeed, the broad band at the binding energy $\leq$\,0.5\,eV universally observed for parent cuprates both for nodal (0,0)-($\pi ,\pi$) and antinodal (0,0)-($0,\pi$) directions\,\cite{Damascelli} can be related with the photoexcitation of the bound electron-hole pair, or EH-dimer. Such a band in the vicinity of the anti-nodal "patches" most likely has nothing to do with any quasiparticle band features such as van Hove singularity\,\cite{Gorkov}. Noticeable dispersion of the "0.5\,eV" band can be attributed to the {\it EH-dimer dispersion rather than to any  
quasiparticle dispersion}. As it was shown by Zhang and Ng\,\cite{Ng} the singlet
two-center CT exciton can move through the antiferromagnetic lattice rather freely in contrast with the single-hole motion. 

Angle-resolved EELS measurements for Sr$_2$CuO$_2$Cl$_2$\,\cite{EELS} point to a noticeable dispersion of the order of 0.2-0.3\,eV for the optically excited \emph{d-d} CT exciton, that agrees with experimentally observed dispersion for the "0.5\,eV" band in ARPES spectra for this and other parent  cuprates\,\cite{Damascelli}.  It is worth noting that a   particularly striking dispersion $\sim$\,0.3\,eV of the \emph{d-d} CT exciton has been revealed by angle resolved EELS for one-dimensional cuprate Sr$_2$CuO$_3$\,\cite{Moskvin2}. Obviously, we should account for different selection rules and matrix element effects for EELS and ARPES.


The relationship between ARPES intensities and the underlying electronic structure can be quite complicated due to matrix element effects\,(see,e.g.Ref.\onlinecite{ARPES}) and caution should be exercised in interpreting detailed features of the ARPES intensities in terms of the spectral function. Nevertheless, the polarization dependent ARPES measurements provide a sensitive test of the symmetries of the excitations with low binding energy.

\section{Evolution of cuprates with non-isovalent substitution}

In contrast with BaBiO$_3$ system where we deal with a  spontaneous generation
of self-trapped CT excitons in the ground state,   the parent insulating
cuprates are believed to be near excitonic instability when the self-trapped CT
excitons  form the candidate relaxed excited states to struggle with the
conventional ground state\,\cite{Toyozawa}. In other words, the lattice relaxed
CT excited state should be treated on an equal footing with the ground state.
Hence,  cuprates are believed to be unconventional systems which are unstable with
regard to a self-trapping  of the low-energy charge transfer excitons with 
nucleation of electron-hole  droplets being actually the system of coupled
electron CuO$_{4}^{7-}$ and hole CuO$_{4}^{5-}$ centers having been glued in
lattice due to strong electron-lattice polarization effects.

What is the evolution of the CuO$_2$ planes in the CT unstable cuprates under a nonisovalent doping?
To describe the evolution  we do start with a very simple model\,\cite{FNT-07} which implies a quantum charge degree of freedom to be the only essential for the cuprate physics. 
We assume only three actual charge states of the CuO$_4$ plaquette: a bare center $M^0$=CuO$_4^{6-}$, a hole center $M^{+1}$=CuO$_4^{5-}$, and an electron center $M^{-1}$=CuO$_4^{7-}$, respectively, forming the charge (isospin) triplet. The system of  such charge  triplets can  be described in frames  of the S=1 pseudo-spin formalism.
  To this end we associate three charge states of the $M$-center with different valence 
  $M^0,M^{\pm}$  with three components of  $S=1$ pseudo-spin (isospin)
triplet with  $M_S =0,\pm 1$, respectively.
  Complete set of the
non-trivial pseudo-spin operators would include three spin-linear (dipole) operators $\hat S_{1,2,3}$ and five independent
spin-quadrupole operators $\{{\hat S_{i}},{\hat S_{j}}\}-\frac{2}{3} {\hat
{\bf  S}}^{2}\delta _{ij}$. Accordingly, to describe different types of  pseudo-spin ordering in such a mixed-valence system
we have to introduce eight order parameters: two classical $diagonal$ order parameters
$\langle {\hat S}_{z}\rangle$ and $\langle {\hat S}_{z}^2\rangle$, and six {\it off-diagonal}
order parameters $\langle {\hat S}_{\pm}\rangle$, $\langle {\hat S}_{\pm}^2\rangle$, and $\langle {\hat T}_{\pm}\rangle$, where $ {\hat T}_{\pm}= ({\hat S}_z {\hat S}_{\pm}+ {\hat S}_{\pm}{\hat S}_z)$.
Diagonal order parameter $\langle {\hat S}_z\rangle$ is related with a valence, or charge density with electro-neutrality
constraint $\sum _{i}\langle {\hat S}_{iz}\rangle =\sum _{i}n_i = n$, while $\langle {\hat S}_{z}^{2} \rangle
= n_p$ determines the density of polar centers $M^{\pm}$, or "ionicity".  The {\it
off-diagonal} order parameters describe different types of the valence mixing, in other words, these can change $valence$ and $ionicity$ with a specific     phase ordering for the disproportionation reaction, single-particle transfer, and for the two-particle transfer. 
 
   An effective pseudo-spin
Hamiltonian of the model mixed-valence system can be written as follows
$$
  \hat H =  \sum_{i}  (\Delta _{i}{\hat S}_{iz}^2
  - h_{i}{\hat S}_{iz}) + \sum_{<i,j>} V_{ij}{\hat S}_{iz}{\hat S}_{jz}+
$$
$$
\sum_{<i,j>} [D_{ij}^{(1)}({\hat S}_{i+}{\hat S}_{j-}+{\hat S}_{i-}{\hat S}_{j+})+ 
D_{ij}^{(2)}({\hat T}_{i+}{\hat T}_{j-}+{\hat T}_{i-}{\hat T}_{j+})]
$$
\begin{equation}
  +\sum_{<i,j>} t_{ij}({\hat S}_{i+}^{2}{\hat S}_{j-}^{2}+{\hat S}_{i-}^{2}{\hat S}_{j+}^{2})\, .
  \label{H}
  \end{equation}
 Two first single-ion terms describe the effects of a bare pseudo-spin splitting,
or the local energy of $M^{0,\pm}$ centers. The second term   may be
associated with a   pseudo-magnetic field $h_i$, in particular, a real electric field. It is easy to see that it describes an electron/hole assymetry. The third term describes the effects of short-  and long-range inter-ionic interactions including screened Coulomb and covalent coupling. 
The last three terms in (\ref{H}) representing the one- and two-particle hopping, respectively, are of primary importance for the transport properties, and deserve special interest. 
Two types of one-particle hopping are
governed by two transfer integrals $D^{(1,2)}$, respectively.
 The transfer integral  $t_{ij}^{\prime}=(D_{ij}^{(1)}+ D_{ij}^{(2)})$
 specifies the  probability amplitude for a {\it local disproportionation, or the $eh$-pair creation}:
$
M^0 +M^0 \rightarrow M^{\pm}+M^{\mp};
$
and the inverse process of the  EH-{\it pair recombination}:
$
M^{\pm}+M^{\mp}\rightarrow M^0 +M^0 ,
$
while the transfer integral  $t_{ij}^{\prime\prime}=(D_{ij}^{(1)}- D_{ij}^{(2)})$
 specifies the  probability amplitude for   a polar center transfer:
 $
 M^{\pm} + M^0 \rightarrow M^0 +M^{\pm},
 $
 or the {\it  motion of the electron (hole) center in the matrix of
$M^0$-centers} or motion of the $M^0$-center in the matrix of $M^{\pm}$-centers.
It should be noted that, if $t_{ij}^{\prime\prime}=0$ but $t_{ij}^{\prime}\not=0$, the EH-pair is locked in a two-site configuration.  At variance with  simple Hubbard-like models where all the types of one-electron(hole) transport are governed by the same
 transfer integral: $t_{ij}^{\prime}=t_{ij}^{\prime\prime}=t_{ij}$, we deal with a "correlated"  single particle transport.
The two-electron(hole) hopping is governed by a transfer integral
  $t_{ij}$, or  a probability amplitude for the exchange reaction:
 $
M^{\pm}+M^{\mp}\rightarrow M^{\mp} +M^{\pm}\,,
$
 or the {\it   motion of the electron (hole) center in the matrix formed by hole (electron) centers}. Obviously, the both ${\hat S}_{\pm}$ and $ {\hat T}_{\pm}$ operators are fermionic, while  ${\hat S}_{\pm}^{2}$  is a bosonic operator.

Simple uniform 
mean-field   phases of the mixed-valent system  include an insulating monovalent $M^0$-phase (parent cuprate), mixed-valence binary (disproportionated) $M^{\pm}$-phase, and mixed-valence ternary (``under-disproportionated'') $M^{0,\pm}$-phase\,\cite{FNT-07}.

In doped cuprates we deal with
 the electron/hole injection to the insulating
parent phase  due to a nonisovalent substitution as in
La$_{2-x}$Sr$_x$CuO$_{4}$, Nd$_{2-x}$Ce$_x$CuO$_{4}$, or change in oxygen
stoihiometry as in YBa$_2$Cu$_3$O$_{6+x}$ and La$_{2}$CuO$_{4+\delta}$. 
Doping is not only to add charge carriers to the system but also to further reduce the gap between electron and hole bands both in electron- and hole-doped copper oxides\,\cite{Xiang,Ando,Gorkov}.
 Anyway the nonisovalent substitution produces  natural
 centers for the  condensation of the CT excitons and the
{\it inhomogeneous nucleation} of EH droplets. Indeed, the gap $\Delta_{CT}$ for the thermal activation of uncoupled electron and hole centers shows a sudden drop from 0.89 to 0.53\,eV upon doping only 1\% of holes to
the parent insulator La$_{2}$CuO$_4$\,\cite{Ando} (or even to 0.25\,eV\,\cite{Gorkov}).

It means the nonisovalent substitution forms the impurity potential centers with a strong inhomogeneous electric field and reduced or even sign reversed $\Delta_i$ values.
At the very beginning of the nucleation regime in the heavily underdoped cuprates the EH droplet nucleates as a nanoscopic cluster composed of several number of neighboring electron and hole
centers pinned  by disorder potential. 
As one of the remarkable  experimental indications to the formation
     of the EH droplets  notice the zero field copper NMR data in
      Y$_{1-x}$Ca$_x$Ba$_2$Cu$_3$O$_6$\,\cite{Mendels}.
 The nonisovalent substitution in the antiferromagnetic state was accompanied by the anomalous
 decrease in the concentration of the NMR resonating copper nuclei: every Ca$^{2+}$ ion leaves
 out from the NMR about 50 copper ions, that could be related with their disproportionation within
  the EH droplet.

Hence, the nonisovalent substitution shifts the phase equilibrium from the parent insulating state ($M^0$-phase) to the binary disproportionated $M^{\pm}$-phase, or a  system of electron CuO$_{4}^{7-}$ and hole
CuO$_{4}^{5-}$ centers. The  system of strongly correlated electron  and hole
 centers  appears to be equivalent to an unconventional electron-hole
Bose-liquid (EHBL) in contrast with the electron-hole  Fermi-liquid in
conventional semiconductors. A simple model description of such a liquid
implies a system of local singlet (S-) bosons with a charge of $q=2e$ moving in a lattice
  formed by hole centers. In a sense the local boson in our scenario represents an electronic equivalent of Zhang-Rice
singlet, or two-electron configuration $b_{1g}^{2}{}^{1}A_{1g}$.


The doping in cuprates such as La$_{2-x}$Sr$_x$CuO$_{4}$ and Nd$_{2-x}$Ce$_x$CuO$_{4}$
gradually shifts the EHBL state away from half-filing making the concentration of the local S-bosons to be $n_B=0.5-x/2$ (LSCO) or $n_B=0.5+x/2$ (NCCO). Nonetheless, in both hole- and electron-doped cuprates we deal with S-bosons moving on the lattice of the hole centers CuO$_4^{5-}$, that makes the unconventional properties of the hole centers to be common ones for the both types of cuprates. 
It is clear that the EHBL scenario makes the doped cuprates the objects of
$bosonic$ physics. There are numerous experimental evidences that support the
bosonic scenario for doped cuprates\,\cite{ASA}. In this connection, we would
like to draw attention to little-known results of comparative
high-temperature studies of thermoelectric power and conductivity which
unambiguously revealed the charge carriers with $q=2e$, or two-electron(hole)
transport\,\cite{Victor}. The well-known relation $\frac{\partial
\alpha}{\partial \ln \sigma}=const=-\frac{k}{q}$ with $|q|=2|e|$ is fulfilled
with high accuracy in the limit of high temperatures ($\sim$\,700\,$\div$\,1000\,K)
for different cuprates (YBa$_{2}$Cu$_{3}$O$_{6+x}$,
La$_{3}$Ba$_{3}$Cu$_{6}$O$_{14+x}$,
(Nd$_{2/3}$Ce$_{1/3}$)$_{4}$(Ba$_{2/3}$Nd$_{1/3}$)$_{4}$Cu$_{6}$O$_{16+x}$).

The evolution of the EH system under doping is particularly revealed in the infrared response of doped cuprates. The ab-plane optical conductivity of eleven single crystals, belonging to the families
Sr$_{2-x}$CuO$_2$Cl$_2$, Y$_{1-x}$Ca$_x$Ba$_2$Cu$_3$O$_6$, and Bi$_2$Sr$_{2-x}$La$_x$CuO$_6$ has been measured recently for a wide range of hole concentrations $0\,<\,p\,<\,0.18$\,\cite{MIR}. At extreme dilution ($p = 0.005$), a weak  narrow peak is first observed at $\approx$ 0.2\,eV (see also Fig.\,\ref{fig2}), that we assign to a dipole-allowed S-P transition in isolated EH-dimers at the energy $\approx 2|t_B|$. For increasing
doping, that peak broadens into a far-infrared (FIR) band whose peak at $\omega_{FIR}$ softens rapidly with doping and whose low-energy edge sets the insulating gap for the bosonic system developed under doping. The insulator-to-metal transition (IMT) occurs when the softening of the FIR band closes the gap thus evolving into a Drude term. In other words, the IMT in cuprates is driven by a conventional transformation of isolated-EH-dimer levels into a conduction bosonic band at a critical $p_{MIT}$. As the Drude intensity progressively increases with doping the MIR band is no more resolved, though an additional oscillator in the mid infrared is required by all Drude-Lorentz fits to the spectra\,\cite{remark}.

The EH-dimer, or coupled electron-hole pair,  can be viewed as a "negative-$U$" center where $U$, or the recombination energy (the energy of the inverse disproportionation reaction) defines an energy scale of a robustness of the EHBL phase. Corresponding intersite \emph{d-d} CT recombination transition 
$$
CuO_4^{5-}+CuO_4^{7-}\rightarrow CuO_4^{6-}+CuO_4^{6-}
$$
can  be a first candidate for a most effective optical destroy 
 of the EH Bose liquid, in particular, suppression of the boson condensate density ("Cooper pair breaking", or CPB  optical effect\,\cite{Zhao,ELi}).
 It seems to be likely that a  famous 1.5\,eV  peak in the optical spectrum of superconducting 123 system that reveals a fairly  sharp CPB resonance\,\cite{ELi} can be assigned to a EH recombination transition with a minimal  energy.  Rather large energy of such an exciton determines the stability of the EHBL phase with regard to its tranformation to the bare parent insulating phase.

Minimal model of the electron-hole Bose liquid is described by a Hamiltonian of local hard-core (hc) bosons on a lattice which can be written in a standard form as follows\,\cite{bubble}:
\begin{equation}
\smallskip
H_{hc}=-\sum\limits_{i>j}t_{ij}{\hat
P}({\hat B}_{i}^{\dagger}{\hat B}_{j}+{\hat B}_{j}^{\dagger}{\hat B}_{i}){\hat P}
+\sum\limits_{i>j}V_{ij}N_{i}N_{j}-\mu \sum\limits_{i}N_{i},  \label{Bip}
\end{equation}
where ${\hat P}$ is the projection operator which removes double occupancy of
any site. 
Here ${\hat B}^{\dagger}({\hat B})$ are
the Pauli creation (annihilation) operators which are Bose-like commuting for
different sites $[{\hat B}_{i},{\hat B}_{j}^{\dagger}]=0,$ if $i\neq j,$  $[{\hat B}_{i},{\hat B}_{i}^{\dagger}]=1-2N_i$,
$N_i = {\hat B}_{i}^{\dagger}{\hat B}_{i}$; $N$ is a full number of sites. $\mu $  the chemical potential
determined from the condition of fixed full number of bosons $N_{l}=
\sum\limits_{i=1}^{N}\langle N_{i}\rangle $ or concentration $\;n=N_{l}/N\in
[0,1]$. The $t_{ij}$ denotes an effective transfer integral,  $V_{ij}$ is an
intersite interaction between the bosons. 
It is worth noting that near half-filling ($n\approx 1/2$) one might introduce the renormalization $N_i \rightarrow (N_i -1/2)$, or neutralizing background, that immediately provides the particle-hole symmetry.

The model of hard-core bosons with an intersite repulsion is
equivalent to a system of spins $s=1/2$  exposed to an external magnetic field
in the $z$-direction. For the system with neutralizing background we arrive at an effective pseudo-spin Hamiltonian
\begin{equation}
H_{hc}=\sum_{i>j}J^{xy}_{ij}({\hat s}_{i}^{+}{\hat s}_{j}^{-}+{\hat s}_{j}^{+}{\hat s}_{i}^{-})+\sum\limits
_
{i>j}
J^{z}_{ij}{\hat s}_{i}^{z}{\hat s}_{j}^{z}-\mu \sum\limits_{i}{\hat s}_{i}^{z}, \label{spinBG}
\end{equation}
where $J^{xy}_{ij}=2t_{ij}$, $J^{z}_{ij}=V_{ij}$, ${\hat s}^{-}= \frac{1}{\sqrt{2}}{\hat B}_ , {\hat s}^{+}=-\frac{1}{\sqrt{2}}
 {\hat B}^{\dagger}, {\hat s}^{z}=-\frac{1}{2}+{\hat B}_{i}^{\dagger}{\hat B}_{i}$,
${\hat s}^{\pm}=\mp \frac{1}{\sqrt{2}}({\hat s}^x \pm i{\hat s}^y)$.

The model of quantum lattice Bose gas has a long history and has been
suggested initially for conventional superconductors\,\cite{Schafroth} and quantum crystals such as $^{4}$He where
superfluidity coexists with a crystalline
order\,\cite{Matsuda-Tsuneto,Fisher}. Afterwards, the Bose-Hubbard (BH) model
has been studied as a model of the superconductor-insulator
transition in materials with  local bosons, bipolarons, or preformed
Cooper pairs\,\cite{Kubo,RMP}. Two-dimensional BH models have been
addressed as relevant to describe the superconducting films and
Josephson junction arrays. The most recent interest to the system of
hard-core bosons comes from the delightful results on Bose-Einstein
(BE) condensed atomic systems produced by trapping bosonic neutral
atoms in an optical lattice\,\cite{Greiner}.

One of the fundamental hot debated  problems in bosonic physics concerns the
evolution of the charge ordered (CO) ground state of 2D hard-core bosons  with a doping away from half-filling.
Numerous model studies steadily confirmed the emergence of "supersolid" CO+BS phases
with simultaneous diagonal CO and off-diagonal Bose superfluid (BS) long range order.
Quantum Monte-Carlo  (QMC) simulations\,\cite{Batrouni} found two significant features of the
2D hard boson model with a
 screened Coulomb repulsion: the absence of supersolid
phase  at half-filling, and a  growing tendency to phase separation (CO+BS)
upon doping away from half-filling.  Moreover, Batrouni and Scalettar\,\cite{Batrouni} studied quantum phase transitions in the ground state
of the 2D hard-core boson Hamiltonian and have shown  numerically that,
contrary to the generally held belief, the most commonly discussed
"checkerboard" supersolid is thermodynamically unstable and phase separates
into solid and superfluid phases. The physics of the CO+BS phase separation in
Bose-Hubbard model is associated with a rapid increase of the energy of a
homogeneous CO state with doping away from half-filling due to a large
"pseudo-spin-flip" energy cost.
Hence, it appears to be energetically more favorable to "extract" extra bosons (holes) from the CO
state and arrange them into finite clusters with a relatively small number of
particles. Such a droplet scenario is believed  to minimize the long-range
Coulomb repulsion.

 The EH Bose liquid in cuprates evolves from the parent phase through to the nucleation of nanoscopic EH droplets around self-trapped CT excitons. However, the EH Bose liquid itself is unstable with regard to a so-called topological phase separation\,\cite{bubble}.  For instance,  deviation from half-filling in EH Bose liquid of quasi-2D cuprates is accompanied by the formation of multi-center topological defect such as charge order (CO) bubble domain(s) with Bose superfluid (BS) and extra bosons both localized in domain wall(s), or a topological CO+BS  phase separation, rather than an uniform mixed CO+BS supersolid phase. A nanosize model of a simplest  topological defect is suggested earlier in Ref.\,\onlinecite{bubble}. Symmetry of the order parameter distribution in the domain wall appears to be specified only by the sign of the boson transfer integral. Problem of the order parameter associated with the bubble domain is much more complicated than in conventional BCS-like approach due to its multicomponent nature. It is worth noting that in frames of BCS-like scenario the symmetry of the order parameter is strictly defined in a momentum space albeit the discussion of different experimental data has usually been performed with a real-space
distributed order parameter. In fact, the EH Bose liquid represents a system with different symmetry of low-lying excited states and competing order parameters that implies their possible ambiguous manifestation in either properties.
Relative magnitude and symmetry of multi-component order parameter are mainly determined by the sign of the $nn$ and $nnn$ bosonic transfer integrals. In general, the topologically inhomogeneous phase of the hc-boson system away from the half-filling can exhibit the signatures both of $s-,d-$, and even $p-$symmetry of the off-diagonal order. Indeed, 
numerous experimental findings point to a more intricate picture with the symmetry of the superconducting state than it is claimed in the conventional uniform d-wave superconductivity.

 The long-wavelength behavior of the hc-boson system is believed to reveal many properties typical for granular superconductors, CDW materials,  Wigner crystals, and multi-skyrmion system akin in a quantum Hall ferromagnetic state of a 2D electron gas\,\cite{Green,Timm}. With decreasing the temperature we deal with isotropic liquid phase, liquid crystal, and crystallization of the multi-skyrmion system, respectively. Namely these charged topological defects with  concentration proportional to the hole/electron doping $x$, rather than single local bosons can be effective charge carriers in doped cuprates.

Making use of a quantum Monte-Carlo technique we have studied the evolution of the phase state of CuO$_2$ planes in  a model CT unstable cuprate La$_{2-x}$Sr$_x$CuO$_4$\,\cite{Korolev}. Tentative results show that the nonisovalent doping gives rise to a nucleation of the inhomogeneous supersolid CO+BS phase  characterized by a charge  and off-diagonal Bose superfluid order parameters which competition results in a generic T-x phase diagram where a pseudogap temperature $T^*(x)$ points to an onset of the CO ordering and a temperature $T_{\nu}(x)>T_c(x)$ with a dome-like $x$-dependence points to an onset of the long-lived BS fluctuations with all the signatures of a local superconductivity. The transition to a bulk 3D coherent superconducting state corresponds to the percolation threshold among the
locally superconducting regions.

It should be emphasized that the minimal model of the EHBL phase in cuprates does not imply intervention of orbital and spin degrees of freedom. Indeed, the model considers a system of the the spin and orbital singlet ${}^1A_{1g}$   local S-bosons moving on the lattice formed by hole centers with the well isolated spin and orbital singlet Zhang-Rice ${}^1A_{1g}$ ground state.

However, both theoretical considerations  and experimental data point towards a more
complicated nature of the valence hole states in doped cuprates than it is
predicted by simple Zhang-Rice model. Actually, we deal with a competition of
conventional hybrid Cu 3d-O 2p $b_{1g}\propto d_{x^2 -y^2}$ state and purely
oxygen nonbonding state with $a_{2g}$ and $e_{ux,y} \propto p_{x,y}$ symmetry\,\cite{NQR-NMR,FNT-11}. Accordingly, the ground state of such a non-Zhang-Rice hole center CuO$_4^{5-}$ as a cluster analog of Cu$^{3+}$ ion should be described by a complex $^1A_{1g}$-$^{1,3}B_{2g}$-$^{1,3}E_u$ multiplet with several competing charge, orbital, and spin order parameters, both of conventional ones (e.g., spin moment or Ising-like orbital magnetic moment) and unconventional, or hidden ones (e.g.,  intra-plaquette's staggered order of Ising-like oxygen orbital magnetic moment or combined spin-quadrupole ordering). The non-Zhang-Rice hole CuO$_4^{5-}$ centers should be considered as singlet-triplet pseudo-Jahn-Teller (ST-PJT) centers  prone to a strong vibronic coupling. 
Novel state of cuprate matter is characterized by a multicomponent order parameter including charge density, U(1) global phase, electric dipole and quadrupole moments, circular orbital current generated by oxygen holes\,\cite{NQR-NMR,FNT-11}. The non-Zhang-Rice structure of the hole CuO$_4^{5-}$ centers forming a lattice for the local boson motion manifest itself in many unconventional properties of the doped cuprates which are often addressed to be signatures of some mechanism of the high-T$_c$ superconductivity. 

Obviously, the local S-bosons do interact with the lattice of the hole centers both as a simple source of fluctuating electric fields  and in a more complex way, in particular, due to a suppression of the ST-PJT order parameters on the hole center occupied by S-boson, that is on the electron center CuO$_4^{7-}$, characterized by occupied Cu 3d$^{10}$ and O 2p$^6$ orbitals. Role of the ST-PJT, or non-ZR-structure of hole centers in superconductivity seems to be merely negative due to effect of a vibronic reduction of the S-boson transfer integral and according increase of its effective mass\,\cite{isotope}. At the same time, the lattice of the ST-PJT hole centers with its large polarizability does provide an effective screening of the boson-boson repulsion thus promoting high T$_c$'s.
Anyway, we should emphasize a crucial role of electron-lattice polarization effects of the order of 1\,eV in high-T$_c$ superconductivity. Namely these effects are believed to provide  a glue to stabilize the electron-hole structure of the EHBL phase.

\section{Conclusion}

A large body of experimental data points towards a charge transfer instability of parent insulating cuprates to be their unique property. We argue that the true charge transfer gap in these compounds is as small as 0.4-0.5\,eV rather than 1.5-2.0\,eV as usually derived from the optical gap measurements. In fact we deal with a competition of the conventional (3d$^9$) ground state and a charge transfer state with formation of electron-hole dimers which evolves under doping to an unconventional bosonic system. We have attempted to incorporate a broad enough collection of experimental results to demonstrate validity of the main message.
Our conjecture does provide an unified standpoint on the main experimental findings for parent cuprates including linear and nonlinear optical, Raman, photoemission, photoabsorption, and transport properties. 
The model approach suggested is believed to provide a conceptual framework for an in-depth understanding  of physics of
 strongly correlated oxides such as cuprates, manganites, bismuthates, and other systems with
 charge transfer excitonic instability and/or mixed valence. We do not attempt here to provide any strict theoretical analysis. The facts we point to are obtained from the unified analysis of  optical, Raman, ARPES, XPS, and Hall experimental data.

In a sense,  the paper provides a validation of a so-called "disproportionation" scenario in cuprates which was addressed earlier by many authors, however, by now it was not properly developed.

The  RFBR Grant No.  10-02-96032 is acknowledged for financial support.

\end{document}